\documentclass[aps,prx,reprint,floatfix,amsmath,amssymb,superscriptaddress,nofootinbib,longbibliography]{revtex4-1}
\usepackage{graphicx}
\usepackage{color}
\usepackage[colorlinks=true]{hyperref}  
\hypersetup{
    bookmarks=true,         
    unicode=false,          
    pdftoolbar=true,        
    pdfmenubar=true,        
    pdffitwindow=false,     
    pdfstartview={FitH},    
    pdfsubject={},   
    pdfcreator={},   
    pdfproducer={}, 
    pdfkeywords={} {} {}, 
    pdfnewwindow=true,      
    colorlinks=true,       
    linkcolor=magenta, 
    citecolor=blue,        
    filecolor=magenta,      
    urlcolor=blue           
} 
\usepackage{bm}
\usepackage{printlen}
\usepackage{units}

\usepackage{afterpage}

\usepackage{ulem}

\renewcommand{\vec}[1]{\mathbf{#1}}

\renewcommand{\L}{\mathcal{L}}

\newcommand{\calE}{\mathcal{E}}
\newcommand{\eff}{{\rm eff}}
\newcommand{\Sch}{\operatorname{Sch}}
\newcommand{\calD}{\mathcal{D}}
\newcommand{\UU}{\operatorname{U}}

\newcommand{\ak}[1]{{\color{blue} #1}}

\usepackage{nicefrac}

\usepackage{ esint }

\definecolor{darkgreen}{rgb}{0.0, 0.26, 0.15}

\definecolor{rred}{rgb}{0.77, 0.12, 0.23}

\definecolor{orange}{rgb}{1.0, 0.49, 0.0}

\newcommand{\beq}{\begin{equation}}
\newcommand{\eeq}{\end{equation}}
\newcommand{\bea}{\begin{eqnarray}}
\newcommand{\eea}{\end{eqnarray}}
\newcommand{\ba}{\begin{array}{ccc}}
\newcommand{\ea}{\end{array}}
\newcommand{\nn}{\nonumber \\}

\begin{document}

\title{Thermoelectric power of Sachdev-Ye-Kitaev islands: \\ Probing Bekenstein-Hawking entropy in quantum matter experiments}


\author{Alexander Kruchkov} 
\affiliation{Department of Physics, Harvard University, Cambridge, MA 02138, USA}
\author{Aavishkar A. Patel}
\affiliation{Department of Physics, University of California Berkeley, Berkeley, CA 94720, USA}
\author{Philip Kim}
\affiliation{Department of Physics, Harvard University, Cambridge, MA 02138, USA}
\author{Subir Sachdev}
\affiliation{Department of Physics, Harvard University, Cambridge, MA 02138, USA}

\date{\today}

\begin{abstract}
The Sachdev-Ye-Kitaev (SYK) model describes electrons with random and all-to-all interactions, and realizes a many-body state without quasiparticle excitations, and a non-vanishing extensive entropy $S_0$ in the zero temperature limit. Its low energy theory coincides the low energy theory of the near-extremal charged black holes with Bekenstein-Hawking entropy $S_0$. Several mesoscopic experimental configurations realizing SYK quantum dynamics over a significant intermediate temperature scale have been proposed. We investigate quantum thermoelectric transport in such configurations, 
and describe the low temperature crossovers out of SYK criticality into regimes with either Fermi liquid behavior, a Coulomb blockade, or criticality associated with Schwarzian quantum gravity fluctuations. Our results show that 
thermopower measurements can serve as a direct probe for $S_0$. 
\end{abstract}

\maketitle

\section{Introduction} 
\label{sec:intro}

The Sachdev-Ye-Kitaev (SYK) model \cite{Sachdev1993,Kitaev2015} is a  strongly interacting quantum many-body system without quasiparticle excitations that is 
maximally chaotic, nearly conformally invariant, and exactly solvable in the limit of large number of interacting particles. It also provide a simple holographic model of charged black holes with AdS$_2$ horizons, and the low energy theories of black holes and the SYK model coincide both at 
leading \cite{SS10,SS15} and sub-leading order \cite{Kitaev2015,Kitaev:2018wpr,JMDS16,JMDS16b,HV16,KJ16,Moitra:2018jqs,Sachdev:2019bjn}.  

There have been a number of interesting proposals towards realizing mesoscopic strongly-interacting correlated electron systems in the presence of disorder and narrow single-particle bandwidth \cite{Franz17,Alicea17,Chen2018,Can2019}; others have advocated realization by quantum gates \cite{Sonner17}. Here, we focus on the `quantum simulation' point-of-view in the mesoscopic realizations. These have the added benefit of providing insight into the relevance of SYK to the naturally occurring correlated electron systems \cite{Altland2019}. A particularly attractive experimental configuration can be built on the zeroth Landau level in irregular-shaped flakes of graphene \cite{Chen2018}, however our study addresses  a more general case of  SYK islands independently of various  possible physical realizations.

The existing theoretical studies of such mesoscopic `SYK islands' have focused on electrical transport between the SYK island and a normal metal lead \cite{Franz17,Alicea17,Chen2018,Can2019,Gnezdilov2018,Altland2019}. Here we extend these analyses to thermoelectric transport in general. We will show below that the 
thermopower  $\Theta$ of an SYK island offers a direct probe of the entropy per particle. In particular, such measurements should be able to extract a unique feature of the SYK model, its non-vanishing extensive entropy in the low temperature ($T$) limit, $S_0 \ne 0$. And this residual entropy is directly connected to the Bekenstein-Hawking entropy of extremal charged black holes \cite{SS10}.

\begin{figure}[b]
\center{
\includegraphics[width = 0.9\columnwidth]{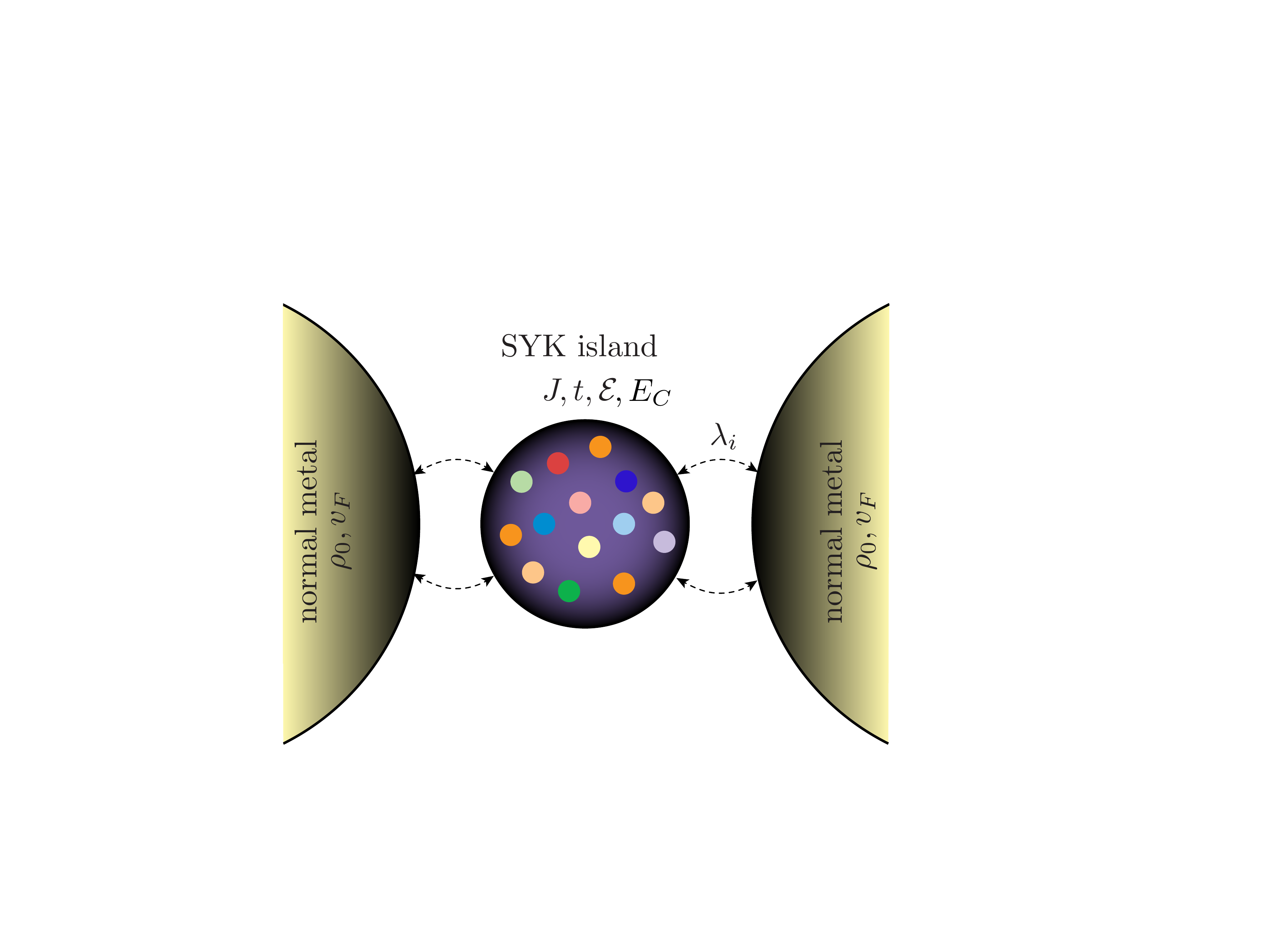} }
\caption{A sketch for an SYK island (center) characterized by a random interaction of mean-square strength $J$, a random electron hopping of mean-square strength $t$,
and a dimensionless particle-hole asymmetry parameter $\mathcal{E}$. The island is coupled to normal metal leads by hopping $\lambda_i$, and this is characterized by an energy scale $\Gamma \propto |\lambda_i|^2 \times$ (density of states in the leads). The SYK behavior requires $\Gamma, t \ll J$.}
\label{fig:sykleads}
\end{figure}
In any realistic experimental configuration, SYK criticality (and the black hole mapping) is only expected to exist in an intermediate temperature scale below the SYK random interaction scale $J$, and above a lower cutoff energy scale. Depending upon the experimental configuration, there are different possibilities for the {\it largest\/} lower energy scale:
\begin{enumerate}
\item The `coherence' scale 
\beq
E_{\rm coh} = \frac{t^2}{J} \,, 
\label{Ec}
\eeq
where $t$ is the bandwidth of the single-particle states. We always assume $t<J$, for otherwise the SYK regime does not exist at any $T$. 
For $T < E_{\rm coh}$, quasiparticles re-emerge, and there is a Fermi liquid regime, whose properties will be recalled in Section~\ref{sec:FL}. The crossover from SYK to Fermi liquid behavior will be described in Section~\ref{sec:SYK}.
\item The charging energy $E_C$. For $T < E_C$, the Coulomb blockade appears, and its interplay with SYK criticality was investigated partly in Ref.~\onlinecite{Altland2019}, but not for the case of thermopower. We will describe this crossover in Section~\ref{sec:charging}.
\item The interaction energy scale $J/N$, where $N$ is the number of single particle states. The crossover at $T \sim J/N$ is associated with quantum gravity fluctuations described by the Schwarzian theory \cite{Kitaev2015,Kitaev:2018wpr,JMDS16,GKST2019}. This crossover has been investigated numerically in Ref.~\onlinecite{Kobrin2020}, but not for transport observables. We will discuss it in Section~\ref{sec:schwarzian}.
\end{enumerate}
Our paper will investigate the associated crossovers out of SYK criticality at these 3 energy scales, especially in the thermopower.

Fig.~\ref{fig:sykleads} illustrates another energy scale, $\Gamma$, associated with the coupling to the metallic leads. All our results in this paper will be in the limit $\Gamma \rightarrow 0$.

\subsection{SYK criticality and thermopower}
\label{sec:sykcriticaliy}

This subsection will recall some of the key properties of the SYK criticality, obtained when all the 4 energy scales just noted are vanishingly small. We also outline our main new results on the thermopower.

Of particular interest to us is how the properties of the complex SYK model (a model built on complex fermions), evolve as a function of the conserved $U(1)$ charge, the electron density $\mathcal{Q}$. 
The ground state of this model realizes a {\it critical phase}, over a range of values of the chemical potential $\mu$, or $\mathcal{Q}$.
For a model with mean-square random interaction of strength $J$, 
the imaginary time electron Green's function obeys at times $|\tau| \gg 1/J$
\beq
G(\tau) \sim \left\{ 
\begin{array}{ccc}
-\tau^{-2\Delta}\, &~& \tau > 0 \\
e^{- 2 \pi \mathcal{E}} (-\tau)^{-2 \Delta} &~& \tau < 0 
\end{array}
\right. \quad, \quad T=0 \label{Gtau}
\eeq
where $\Delta=1/4$ is the scaling dimension of the electron operators. Our interest here focuses particularly on 
the particle-hole asymmetry $\mathcal{E}$, which can be expressed as a function of $\mathcal{Q}$
via a Luttinger-like relation \cite{GPS01,Davison2017,GKST2019}.
\begin{align}
 \mathcal{E} &= \frac{1}{2 \pi} \ln \frac{\sin(\pi \Delta + \theta  )}{\sin(\pi \Delta - \theta  )},  \nonumber \\
\mathcal{Q} &= \frac{1}{2} - \frac{\theta}{\pi} + \left( \Delta - \frac{1}{2} \right) \frac{\sin(2 \theta)}{\sin(2 \pi \Delta)} \,. \label{EQtheta}
\end{align}
For the specific case of the pure SYK model with infinite-range interactions, numerical studies \cite{Sachdev1993,GPS99,GPS01,Fu:2016yrv,Azeyanagi2018}
show that solutions with variable $\mathcal{Q}$ and $\mathcal{E}$ exist for $0.2 < \mathcal{Q} < 0.8$ or $|\mathcal{E}| < 0.14$ \cite{Azeyanagi2018,Patel2019}.

Note that the particle-hole asymmetry in (\ref{Gtau}) is significantly stronger than that in a Fermi liquid case. Even in the presence of an energy dependent density of states, the electron Green's function of a Fermi liquid is particle-hole symmetric, with $G(\tau) \sim - {1}/{\tau}$ at large $|\tau|$, with the same amplitude for both signs of $\tau$. So formally, $\mathcal{E} = 0$ for a Fermi liquid.

The strong particle-hole asymmetry of the SYK model is intimately connected to its extensive entropy as $T \rightarrow 0$ via the relation \cite{GPS01,Davison2017,GKST2019}
\beq
 \frac{d \mathcal{S}}{d \mathcal{Q}} = 2 \pi \mathcal{E} , \label{dSdQ}
\eeq
where $N$ is the number of sites in the SYK model, and $\mathcal{S} \equiv S_0/N$ is entropy density in the limit where $N \rightarrow \infty$ first, followed later by $T \rightarrow 0$. 
(Eq.~(\ref{dSdQ}) also shows that $\mathcal{E}=0$ in a Fermi liquid, because $\mathcal{S}$ vanishes as $T\rightarrow 0$ in a Fermi liquid.)
The relationship (\ref{dSdQ}) was obtained by Georges et al. \cite{GPS01}, building upon large $N$ studies of the multichannel Kondo problem \cite{PGKS98}.
Independently, this relationship appeared as a general property of black holes with AdS$_2$ horizons \cite{Sen05,Sen08}, where $\mathcal{E}$ is identified with the electric field on the horizon \cite{Faulkner09}.

We now turn to the thermopower, $\Theta$, and its connection to the entropy. In Fermi liquids, the thermopower is usually computed by the `Mott formula'
\beq
\Theta = \frac{\pi^2}{3} \frac{k_B^2 T}{e} \frac{ \partial \ln \sigma}{\partial \epsilon_F} , \label{Mott}
\eeq
where $\sigma$ is conductivity at Fermi energy $\epsilon_F$ (we set $k_B =1$ elsewhere). Note that the thermopower vanishes as $T \rightarrow 0$, and is proportional to particle-hole asymmetries which lead to a Fermi energy dependence in the conductivity. If we assume that such dependence is entirely due to the density of states (and not due to the scattering time), then (\ref{Mott}) can be written as
\beq
\Theta = \frac{1}{e} \frac{d \mathcal{S}}{d \mathcal{Q}}, \label{Kelvin}
\eeq
an expression dubbed as the `Kelvin formula' in Ref.~\onlinecite{Peterson2010}. 

 As the entropy vanishes linearly with $T$ in a Fermi liquid, (\ref{Kelvin}) implies that the thermopower also vanishes linearly with $T$. 
The Kelvin formula was originally proposed as an approximate empirical formula useful in certain strongly correlated
systems \cite{Peterson2010,Georges16}.

Turning to thermopower in systems without quasipasiparticles, Davison {\it et al\/} \cite{Davison2017} showed that the Kelvin formula was {\it exact} for transport in lattices of coupled SYK islands. 
Given the non-vanishing of $\mathcal{S}$ as $T \rightarrow 0$, this implies that $\Theta$ also remains non-zero as $T \rightarrow 0$, in striking contrast from a Fermi liquid.
It was also found \cite{Davison2017} that the Kelvin formula was an exact general feature of charged black holes with AdS$_2$ horizons. 
We are interested here in transport between a single SYK island and normal metal leads, a general idea is sketched in Fig.~\ref{fig:sykleads} (in contrast to transport between 2 SYK islands in the lattice models \cite{Gu:2017ohj,Davison2017,Song2017,Zhang2017,Chowdhury2018,Patel2018}). 
One of our new results is that in the single-channel lead configuration, and under conditions in which the transport is dominated by SYK correlations (specified more carefully in the body of the paper), we have the thermopower
\beq
\Theta = \frac{2}{3e} \, 2 \pi \mathcal{E} \label{ThetaE} .
\eeq
So as in other systems, the thermopower is intimately connected to the $\mathcal{Q}$ dependence of the entropy, via (\ref{dSdQ}). 
One of our main observations is that the relations (\ref{Gtau}), (\ref{dSdQ}), and (\ref{ThetaE}) link together the surprising features of the SYK model:
({\it i\/}) the strong particle-hole asymmetry in the low energy limit, ({\it ii\/}) the non-vanishing extensive entropy as $T \rightarrow 0$, and
({\it iii\/}) the non-vanishing  thermoelectric power as $T \rightarrow 0$. 

The outline of the remainder of the paper is as follows. We will setup the basic formalism for the transport across SYK islands in Section~\ref{sec:setup}. To set the stage, Section~\ref{sec:FL} will investigate the familiar disordered Fermi liquid regime, when $E_{\rm coh}$ is the largest of the low energy cutoffs, and $T< E_{\rm coh}$. Then Sections~\ref{sec:SYK}, \ref{sec:charging}, and \ref{sec:schwarzian} will discuss the low $T$ crossover out of SYK criticality controlled by the 3 energy scales noted earlier in this introduction. We will discuss possible values of experimental parameters, and these energy scales in Section~\ref{sec:expts}.

\section{Setup}
\label{sec:setup}

We will model the SYK island by a Hamiltonian, $H_{\text{I}}$ with random interactions and hopping, and a bare charging energy $E_C^0$:
\bea
&~& H_{\text{I}} = \frac{1}{(2 N)^{3/2}} \sum_{ij;kl = 1}^N J_{ij;kl} c^{\dag}_i c^{\dag}_j c^{}_k c^{}_l  \\
&~&+ \frac{1}{N^{1/2}} \sum_{ij=1}^N t_{ij} c^{\dag}_i c^{}_j
- \mu \sum_i c^{\dag}_i c^{}_i + \frac{E_C^0}{2} \left( \sum_i c^{\dag}_i c^{}_i \right)^2 \,. \nonumber
\label{Hamiltonian}
\eea
Here $J_{ij;kl}$ is random interaction with zero mean and root-mean-square magnitude $J$ ($\langle J_{ij;kl} \rangle= 0$, $\langle |J_{ij;kl}|^2 \rangle=J^2$), and $t_{ij}$ is random hopping with zero mean and root-mean-square magnitude $t$ ($\langle t_{ij} \rangle= 0$, $\langle |t_{ij}|^2 \rangle=t^2$). The bare charging energy $E_C^0$ is renormalized to $E_C$ by the $J_{ij;kl}$ interactions, as was computed in Ref.~\onlinecite{GKST2019}; we will use the renormalized $E_C$ in all our results. 

The large $N$ retarded electron Green's function, $\mathcal{G} (\omega)$ can be computed numerically at $E_C =0$ for all $\omega$, $T$, $\mu$, $J$, and $t$
by solving a set of integro-differential equations. The solutions of these equations where described elsewhere in the literature \cite{OPAG99,Song2017} for $\mathcal{E}=0$. We have extended these numerics to non-zero $\mathcal{E} (\mathcal Q) $,  with 
\beq
\mathcal Q = \frac{1}{N} \sum_i \langle c^{\dag}_i c^{}_i \rangle\,, 
\eeq
and will describe the implications for thermoelectric transport in Section~\ref{crossoverFL}. We will also describe the extension to non-zero $E_C$ (at $t=0$) in Section~\ref{sec:charging}, and the Schwarzian fluctuations at non-zero $J/N$ in Section~\ref{sec:schwarzian}.

A crucial feature of this solution for $t \ll J$ (which we assume throughout) is the emergence of a low energy scale $E_{\rm coh}$ which was defined in (\ref{Ec}) as the first of low energy cutoffs for SYK behavior. When $E_{\rm coh}$ is the largest of the low energy cutoffs, then for $T<E_{\rm coh}$ (but $T$ larger than the other of the low energy cutoffs),
we recover the physics of a disordered Fermi liquid, albeit with strong renormalization from the interactions. 

Let us now couple the island to the leads (Fig.~\ref{fig:sykleads}). Here we follow the approach of Gnezdilov {\it et al.} \cite{Gnezdilov2018}, and model the leads by
\begin{align}
\mathcal{H}  = H_{\text{I}} + \sum_{\vec q} \varepsilon_{\vec q} \, a^{\dag}_{\vec q} a^{}_{\vec q} 
+ 
\sum_{i, \vec q} \lambda_i c^{\dag}_{i}  a_{\vec q} + \lambda^*_i a^{\dag}_{\vec q} c^{}_{i} .
\label{Hamiltonian1}
\end{align}
The new terms represent  dispersive electrons 
($a^{\dag}_{\vec p}$, $a^{}_{\vec p}$) in the single-channel contact with a  dispersion  $\varepsilon_{\vec{q}} $ near the Fermi surface. The SYK quantum dot fermions $c^{\dag}_i$ and the lead  fermions with $a_{\vec p}$ are coupled by the random hopping $\lambda_i$. The coupling to the leads will be characterized by the energy scale 
\beq
\Gamma = \pi \rho_{\rm lead} \sum_i |\lambda_i|^2 ,
\eeq
where $\rho_{\rm lead}$ is the density of states in the lead. 
In the limit $\Gamma \rightarrow 0$, the non-linear electrical transport of $\mathcal{H}$ is described by the differential conductance \cite{Gnezdilov2018,Costi2010}
\beq
\frac{dI}{dV} = \frac{4 \Gamma e^2}{h} \int_{-\infty}^{\infty} d \omega f'(\omega - eV)  \mbox{Im} \mathcal{G} (\omega) \label{dIdV}
\eeq
where $\mathcal{G}$ is the retarded Green's function of $H_I$ alone, $f(\omega)=1/(e^{\omega/T}+1)$ is the Fermi function, and a factor of 2 for electron spin has been included. 

For linear thermoelectric transport, we use expressions derived by Costi and Zlati\'c \cite{Costi2010}.  
We define 
\begin{align}
\L_{ij} =    \frac{2 \Gamma}{\pi \hbar } \int \limits_{-\infty}^{+\infty}  d \omega \,  \omega ^{i+j -2} f'(\omega) \text{Im} \mathcal G (\omega)\,.
\label{Ons-transport}
\end{align}
Then we have for the electric conductance $\sigma$, thermal conductance $\kappa$, and the thermopower $\Theta$ (see also  Ref.~\onlinecite{Mahan,Xu2011}) 
\bea
\sigma &=&  e^2 \,  \L_{11}, \nn
\kappa &=& \beta \left(\L_{22} - \frac{\L_{12}^2}{\L_{11}} \right) , \nn 
\Theta &=& \frac{\beta}{e} \frac{\L_{12}}{\L_{11}} ,
\eea
where $\beta = 1/T$. The following sections will evaluate these expressions in different regimes depending upon the relative values of the small energy scales
$T$, $E_{\rm coh}$, and $E_C$.

\section{Crossover from SYK to Fermi Liquid}
\label{crossoverFL}

We begin with the case in which $E_{\rm coh}$ in (\ref{Ec}) is the largest of low energy cutoff scales {\it i.e.\/} $E_{\rm coh}, T \gg J/N, E_C$. Then there is a crossover from SYK criticality for $E_{\rm coh} \ll T \ll J$ to a Fermi liquid regime for $E_C \ll T \ll E_{\rm coh}$. 

\begin{figure}
\center{
\includegraphics[width = 2.9in]{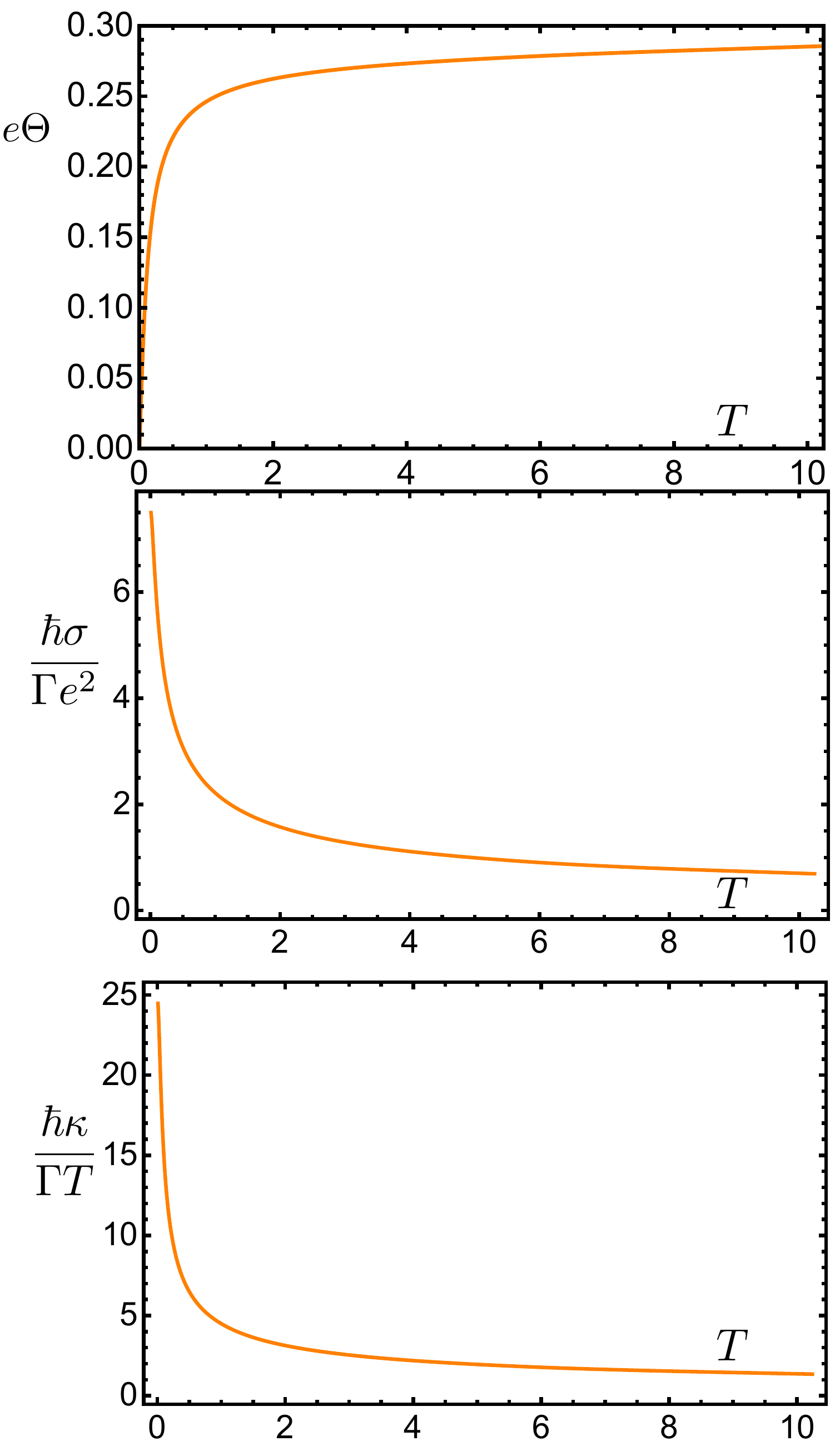} }
\caption{Crossovers in transport from the SYK regime at $T > E_{\rm coh}$, to Fermi liquid behavior at $T < E_{\rm coh}$. There is no divergence in the conductances as $T \rightarrow 0$, and they saturate at values related to those in (\ref{FermiLiquidregime}), with an energy scale of order $E_{\rm coh}$ replacing $t,\mu$. It is assumed here that $T, E_{\rm coh} \gg E_C, J/N$. We use $J=10$, $t=0.1$, and $\mathcal{Q}=1/3$.
}
\label{fig:FL_crossover}
\end{figure}

A full description of the crossover requires 
numerical results for $\mathcal{G} (\omega)$, which generalize the earlier computations \cite{OPAG99,Song2017} to non-zero $\mathcal{E}$. 
Our numerical results for crossover bewteen these two regimes at fixed filling $\mathcal Q$  are presented in Fig.~\ref{fig:FL_crossover}.

More complete analytic results are possible for the limiting SYK regime for $E_{\rm coh} \ll T \ll J$, and for the Fermi liquid regime for $T \ll E_{\rm coh}$, and we will present them in the following subsections.

\subsection{Pure SYK regime}
\label{sec:SYK}

We consider the main regime of interest to us, when $T$ is larger than all the low energy cutoffs mentioned in Section~\ref{sec:intro}, but we have $T \ll J$.

To leading order, we may set $E_{\rm coh}=0$. Analytic expressions for the Green's function are possible for $t/J=0$ and $T \ll J$. Then we can use the conformal form of the (retarded) SYK Green's function
\begin{align}
\mathcal{G} (\omega)  = & - i e^{- i \theta} \left( \frac{\pi} {\cos 2 \theta} \right) ^{1/4} \left( \frac{\beta}{2 \pi J} \right)^{1/2} 
\nonumber
\\
& \times
\frac{\boldsymbol \Gamma({\frac{1}{4} - \frac{i}{2 \pi}\beta \omega} + i \mathcal{E} )}{\boldsymbol \Gamma({\frac{3}{4} - \frac{i}{2 \pi}\beta \omega} + i \mathcal{E})} , 
\label{G_SYK}
\end{align}
where $\boldsymbol \Gamma (x)$ is the Gamma function. 

We find that the SYK state has a finite, temperature-independent thermopower, a consequence of finite Bekenstein-Hawking entropy $S_0$ associated with the SYK state.  In the case of single-island pure SYK flake we find the  thermopower $\Theta$ is directly connected to particle-hole asymmetry $\mathcal E$:
\begin{align}
\Theta  = \frac{\beta}{e}
\frac{ 
\int_{-\infty}^{+\infty} d\omega \, \omega f'( \omega  )
\,\mbox{Im} \mathcal{G} (\omega) 
}
{
\int_{-\infty}^{+\infty} d\omega  f'( \omega )
\,\mbox{Im} \mathcal{G} (\omega) 
} 
 = \frac{4 \pi}{3e} \mathcal{E}  . 
 \label{thermopower-E}
\end{align}
This $T$-independent thermopower holds for $T>E_{\rm coh} \sim t^2/J$, and vanishes linearly with $T$ in the Fermi liquid regime $T < E_{\rm coh}$ (we are assuming here that the other cutoff energy scales discussed in Section~\ref{sec:intro} are even smaller than $E_{\rm coh}$

It is useful to compare (\ref{thermopower-E}) with the result obtained for the case of thermoelectric transport between SYK islands, in contrast to the transport with normal metal leads, as in Fig.~\ref{fig:sykleads}. Then the corresponding expression for the thermopower was obtained by
Patel {\it et al.} \cite{Patel2018}
\begin{align}
\Theta   = \frac{\beta}{e}  , 
\frac{ 
\int_{-\infty}^{+\infty} d\omega \, \omega f'( \omega  )
\,[\mbox{Im} \mathcal{G} (\omega) ]^2
}
{
\int_{-\infty}^{+\infty} d\omega  f'( \omega )
\, [\mbox{Im} \mathcal{G} (\omega) ]^2
} 
 = \frac{2 \pi}{e} \mathcal{E}  . 
 \label{thermopower-E2}
\end{align}
Note that we now have 2 powers of the Green's function in the numerator and denominator because both the initial and final locations of the electron are on SYK islands. Moreover, we find that the Kelvin relation (\ref{Kelvin}) is then precisely satisfied.

The remaining transport properties in the SYK regime can be obtained by inserting (\ref{G_SYK}) into the expressions in Section~\ref{sec:setup}. We obtain for $T\ll J$, and $T \gg E_{\rm coh}, J/N, E_C$:
\bea
e \Theta &=& \frac{4 \pi}{3} \ak {\mathcal E}, \nonumber \\
\frac{\hbar \sigma}{\Gamma e^2} &=& \left[ 0.72 + \mathcal{O}(\calE^2) \right] \frac{1}{\sqrt{JT}} \nonumber \\
\frac{\hbar \kappa}{\Gamma T} &=&  \left[ 1.42 + \mathcal{O}(\calE^2) \right] \frac{1}{\sqrt{JT}} 
\label{SYKregime}
\eea
(the exact value of conductance prefactor is $2 \sqrt{2} \pi^{-1/4} \Gamma(3/4)/\Gamma(1/4) = 0.72$). The Lorenz ratio linking electrical and thermal conductance is
\begin{align}
\frac{e^2 \kappa}{\sigma T} = \frac{\pi^2}{5} + \frac{1}{5} \left(\frac{4 \pi}{3} \mathcal E \right)^2,  
\end{align}
and is $T$-independent, but its value is distinct from the Lorenz number $= \pi^2/3 \approx 3.29$ in a Fermi liquid.

Recently, themoelectric cooling and power generation in nanodevices, characterized by computed thermoelectric figure of merit, has attracted significant attention \cite{Roura2018,Sheng2019,Fu2019}. We calculate this quantity for a SYK island. The maximum efficiency of electricity generation by a thermoelectric device is typically described in terms of its  figure of merit  $Z$:
\begin{align}
Z = \frac{  \mathcal L_{12}^2}{\det \mathcal L_{i j}} = \frac{\Theta^2  \sigma T}{\kappa}\,.
\end{align}
For SYK criticality, thethermoelectric figure of merit is
\begin{align}
Z  = \frac{5 (e  \Theta)^2}{\pi^2 + (e \Theta) ^2}\,.
\end{align}
This can be larger than in a Fermi liquid which has $Z \sim (T/E_F)^2$, where $E_F$ is the Fermi energy.

\subsection{Fermi liquid regime}
\label{sec:FL}

Next, we present the familiar results in the Fermi liquid regime $T \ll E_{\rm coh}$, 
but $T \gg E_C$.
The random SYK interactions are formally irrelevant, 
and can be accounted for by a renormalization in the value of $t$ to $E_{\rm coh}$ \cite{OPAG99,Song2017}. We just present the results without this renormalization here,
but it should be kept in mind that for the results in this section $t$ should be replaced by an energy of order $E_{\rm coh}$ when $t \ll J$.

The Green's function can be computed exactly in the absence of interactions by solving a quadratic equation \cite{OPAG99}
\beq
\mathcal{G} (\omega) = \frac{1}{2t^2} \left( \omega + \mu - \sqrt{(\omega+\mu)^2 - 4 t^2} \right)\, ,
\eeq
which yields the well-known semi-circular density of states of a random matrix. 
We denote 
\bea
\rho &=& - \frac{1}{\pi} \mbox{Im} \mathcal{G} (0) \nonumber \\
\rho' &=& - \frac{1}{\pi} \left. \frac{d}{d\omega} \mbox{Im} \mathcal{G} (\omega)\right|_{\omega =0}\,,
\eea
as the density of states, and its derivative, at the Fermi level. 
Then the thermoelectric responses are 
\bea
e\Theta &=& \frac{\pi^2 T}{3 } \frac{\rho'}{\rho} \nonumber \\
\frac{\hbar \sigma}{\Gamma e^2} &=& 2    \rho   \nn
\frac{\hbar \kappa}{\Gamma T} &=& \frac{2\pi^2 }{3 } \rho \,.
\label{FermiLiquidregime}
\eea
As we can see, the thermopower $\Theta$ vanishes as $T \to 0$, which is a general phenomenon for interacting matter with quasiparticles. Both the Mott Law and Wiedemann-Franz Law hold, and the Lorentz number is $L = e^2 \kappa/(\sigma T) = {\pi^2}/{3} $, as well as the Mott number is $M = \Theta/ (T \rho '/ (e \rho)) = {\pi^2}/{3} $.

\section{Charging energy}
\label{sec:charging}

We now turn to the case where the charging energy $E_C$ is the largest of the low energy cutoffs. We will consider the crossover around $T \sim E_C$, with $T < J$, as always. But in this section we assume $E_C \gg E_{\rm coh}, J/N$.

Here we have consider the consequences of fluctuations in the total charge, $Q = N \mathcal{Q}$, of the SYK ideal. A complete theory for such fluctuations was presented recently in Ref.~\onlinecite{GKST2019}, and we recall essential aspects for our purposes.

We need the corrections to the SYK Green's function from the charge fluctuations. This correction will arise from the fluctuations of a conjugate phase field $\lambda (\tau)$, and a time reparameterization field $\varphi (\tau)$.
From Eqs (2.47) and (2.48) of Ref.~\onlinecite{GKST2019}, the Green's function of the complex SYK model is given by
\bea
&~& \mathcal{G}(\tau_1-\tau_2) = - b^\Delta \biggl\langle \varphi' (\tau_1)^{\Delta} \varphi' (\tau_2)^{\Delta} \label{G1} \\
 &~&~~~~~~~~~~~~~~~\times  \left( 2 \sin \frac{\varphi(\tau_1) - \varphi (\tau_2)}{2} \right)^{-2\Delta} \nonumber \\
 &~&\times \exp \Bigl[ i \lambda(\tau_1)-i \lambda(\tau_2)+\calE \left( \pi - \varphi(\tau_1) + \varphi (\tau_2)  \right) \Bigr] \biggr\rangle_{I_{\eff} [\varphi, \lambda]} \nonumber 
\eea
where the imaginary action for the fields $\varphi (\tau)$ and $\lambda(\tau)$ is given by
\bea
 I_{\eff} [\varphi, \lambda]
&=& -S_0 (Q) + \frac{1}{2E_C} \int_0^{\beta} d \tau
\bigl(\lambda'(\tau) + i\calE\varphi'(\tau)\bigr)^2 \nonumber \\
&-&  C_{\text{Sch}} \int_0^{\beta} d \tau \,
\Sch\left( \tan\frac{\varphi(\tau)}{2},\, \tau\right) \,.
\label{Seff}
\eea
Here $\Delta = 1/4$ is the scaling dimension of the electron operator, the prefactor $b$ is 
\beq
b = \frac{(1-2 \Delta)\sin(2 \pi \Delta)}{ 2 \pi J^2 (\cosh(2 \pi \calE) + \cos(2 \pi \Delta))} \,,
\eeq
$S_0 (Q)$ is the zero temperature entropy at the equilibrium total charge $Q$, the field $\varphi(\tau)$ is a
monotonic time reparameterization obeying 
\beq
\varphi(\tau + \beta) = \varphi(\tau) + 2\pi\,, \label{phiperiod}
\eeq
and $\lambda (\tau)$ is a phase field  obeying 
\beq
\lambda (\tau + \beta) = \lambda (\tau) + 2 \pi n\,, \label{lambdaperiod}
\eeq
with integer winding number $n$ conjugate to the total charge $Q$.  The notation 
$\Sch ( f(x), x )$ stands for the Schwarzian derivative
\begin{equation}
\Sch ( f(x), x ) := \frac{f'''}{f'} - \frac{3}{2} \left( \frac{f''}{f'} \right)^2 \,.
\end{equation}
In (\ref{Seff}), we have replaced the coupling $NK \sim N/J$ in Ref.~\onlinecite{GKST2019} by the inverse of the charging energy, $1/E_C$, to correspond to the notation of Ref.~\onlinecite{Altland2019}, and the co-efficient of the Schwarzian, $N \gamma/(4 \pi^2) \sim N/J$, by $C_{\text{Sch}}$ to correspond to the notation of Ref.~\onlinecite{Kobrin2020}.

In the leading large $N$ limit, we can replace $\varphi (\tau)$ and $\lambda (\tau)$ in (\ref{G1}) by their saddle-point values $\varphi (\tau) = 2\pi \tau/\beta$ and $\lambda (\tau) = 0$. This yields a $\mathcal{G}(\tau)$ whose Fourier transform is (\ref{G_SYK}).

We now turn to fluctuations. In the present section, with $E_C \gg J/N$, the fluctuations of $\lambda (\tau)$ are more important than those $\varphi (\tau)$.
As the action for $\lambda (\tau)$ is Gaussian, the path integral over $\lambda$ is relatively easy to evaluate.
First we consider the partition function
\begin{equation}
\frac{\mathcal{Z}_\lambda}{e^{S_0 (Q)}} =  \int \frac{\calD \lambda }{\UU(1)} \exp \left( -\frac{1}{2E_C} \int_0^{\beta} d \tau
\bigl(\lambda'(\tau) + i\calE\varphi'(\tau)\bigr)^2 \right) \,, \nonumber
\end{equation}
where we have divided the integral by the volume of $\UU(1)$ because we should view $\UU(1)$ as a gauge symmetry. The prefactor $e^{S_0 (Q)}$ reflects the multiplicity of the states. We will see that this partition function and the correlator needed for (\ref{G1}) are independent of $\varphi (\tau)$.
Related partition functions have also been evaluated recently in Ref.~\onlinecite{Liu:2019niv}.
We now change variables to $\tilde{\lambda} (\tau)$, where
\beq
\lambda(\tau) =\tilde{\lambda}(\tau) + 2 \pi n \frac{\tau}{\beta} +i\calE \left( \frac{2\pi}{\beta}\tau-\varphi(\tau) \right)\,; \label{tlambda}
\eeq
from (\ref{phiperiod}) and (\ref{lambdaperiod}) we see that  $\tilde{\lambda}(\tau)$ is periodic
\beq
\tilde{\lambda} (\tau + \beta) = \tilde{\lambda} (\tau)\,.
\eeq
Then we can write the path integral over $\tilde{\lambda}$ as
\bea
\mathcal{Z}_\lambda &=& e^{S_0 (Q)} \left( \, \sum_{n=-\infty}^{\infty} \exp \left[ - \frac{2 \pi^2 }{\beta E_C} (n + i \calE)^2 \right]  \right) \nonumber \\
&\times& \int \frac{\calD \tilde{\lambda} }{\UU(1)} \exp \left[- \frac{1}{2E_C} \int_0^{\beta} d \tau \left(\tilde{\lambda}'(\tau)  \right)^2 \right] \,. \label{ZQ1}
\eea
The second term is just the imaginary time amplitude for a `free particle' of mass $1/E_C$, moving on an infinite line, to return to its starting point in a time $\beta$ \cite{ssbook}. In this manner, we obtain
\begin{eqnarray}
\mathcal{Z}_\lambda &=& \sqrt{\frac{2 \pi}{\beta E_C}} e^{S_0 (Q)}  \, \sum_{n=-\infty}^{\infty} \exp \left[ - \frac{2 \pi^2 }{\beta E_C} (n + i \calE)^2 \right]\,. \label{ZQ3}
\end{eqnarray} 
This expression is convergent for all values of $\beta E_C$, but rapidly so when $\beta E_C \ll 1$, when the leading contribution is just the $n=0$ term. Conversely, when $\beta E_C \gg 1$, it is easier to use the equivalent expression obtained from the Poisson summation formula
\begin{eqnarray}
\mathcal{Z}_\lambda &=& e^{S_0 (Q)}   \, \sum_{p=-\infty}^{\infty} \exp \left[ - \frac{\beta E_C p^2}{2} + 2 \pi \calE p \right] \nonumber \\
&=&  \sum_{p=-\infty}^{\infty} e^{S_0 (Q+p)} \exp \left[ - \frac{\beta E_C p^2}{2} \right]\,, \label{ZQ2}
\end{eqnarray} 
where we have used (\ref{dSdQ}).
The expression (\ref{ZQ2}) has a simple interpretation: it is the sum over states with total charge $Q+p$,
with near-degeneracy $e^{S_0 (Q+p)}$, and energy $(1/2)E_C p^2$. For $\beta E_C \gg 1$, the $p=0$ term dominates.

Next, we consider the Green's function
\bea
&& \mathcal{G}_\lambda (\tau_1 - \tau_2) =  \\
&&~\biggl\langle \exp \Bigl[ i \lambda(\tau_1)-i \lambda(\tau_2)+\calE \left( \pi - \varphi(\tau_1) + \varphi (\tau_2)  \right) \Bigr] \biggr\rangle_{\mathcal{Z}_\lambda}\,.\nonumber
\eea
Evaluating this with the parameterization (\ref{tlambda}), in a manner similar to $\mathcal{Z}_\lambda$, we obtain
\bea
\mathcal{G}_\lambda (\tau) &=&  \exp \left[ \calE \left( \pi - \frac{2 \pi}{\beta} \tau  \right) \right] 
\frac{e^{S_0 (Q)}}{ \mathcal{Z}_\lambda} \sqrt{\frac{2 \pi}{\beta E_C }}   \nonumber \\
&~&~~~\times \sum_{n=-\infty}^{\infty} \exp \Biggl[i \frac{2 \pi n}{\beta} \tau   \\
&~&~~~~ - \frac{2 \pi^2 }{\beta E_C} (n + i \calE)^2 
- \frac{E_C}{\beta} \sum_{\omega_m \neq 0} \frac{1 - e^{i \omega_m \tau}}{\omega_m^2} \Biggr]\,,
\nonumber
\eea
where $\omega_m$ is a bosonic Matsubara frequency. The summation over $\omega_m$ can be evaluated exactly \cite{Kamenev96}, and we obtain (assuming henceforth that $|\tau| \leq \beta$)
\bea
&& \mathcal{G}_\lambda (\tau) =  \exp \left[ \calE \left( \pi - \frac{2 \pi}{\beta} \tau  \right) - \frac{E_C}{2} \left( |\tau| - \frac{\tau^2}{\beta} \right) \right] 
\label{GL2}  \\
&& \times \frac{e^{S_0 (Q)}}{ \mathcal{Z}_\lambda} \sqrt{\frac{2 \pi}{\beta E_C}} \sum_{n=-\infty}^{\infty} \exp \left[i \frac{2 \pi n}{\beta} \tau - \frac{2 \pi^2 }{\beta E_C} (n + i \calE)^2 
  \right]\,. \nonumber
\eea
This expression is rapidly convergent when $\beta E_C \ll 1$, and the leading answer is the $n=0$ term. For $\beta E_C \gg 1$, it is easier to use the equivalent expression obtained from the Poisson summation formula
\bea
&~& \mathcal{G}_\lambda (\tau) =  \frac{e^{S_0 (Q)}}{ \mathcal{Z}_\lambda} \exp \left[ \pi \calE - \frac{E_C}{2}  |\tau| \right] \nonumber \\
&~&\times \sum_{p =-\infty}^{\infty} \exp \left[ - \frac{\beta E_C p^2}{2} + 2 \pi p \left(\calE  - \frac{E_C \tau}{2 \pi} \right)  \right]\,. \label{GL1}
\eea
The expression (\ref{GL1}) agrees with Eq. (11) in the supplement of Ref.~\onlinecite{Altland2019}.

\begin{figure}
\center{
\includegraphics[width = 2.9in]{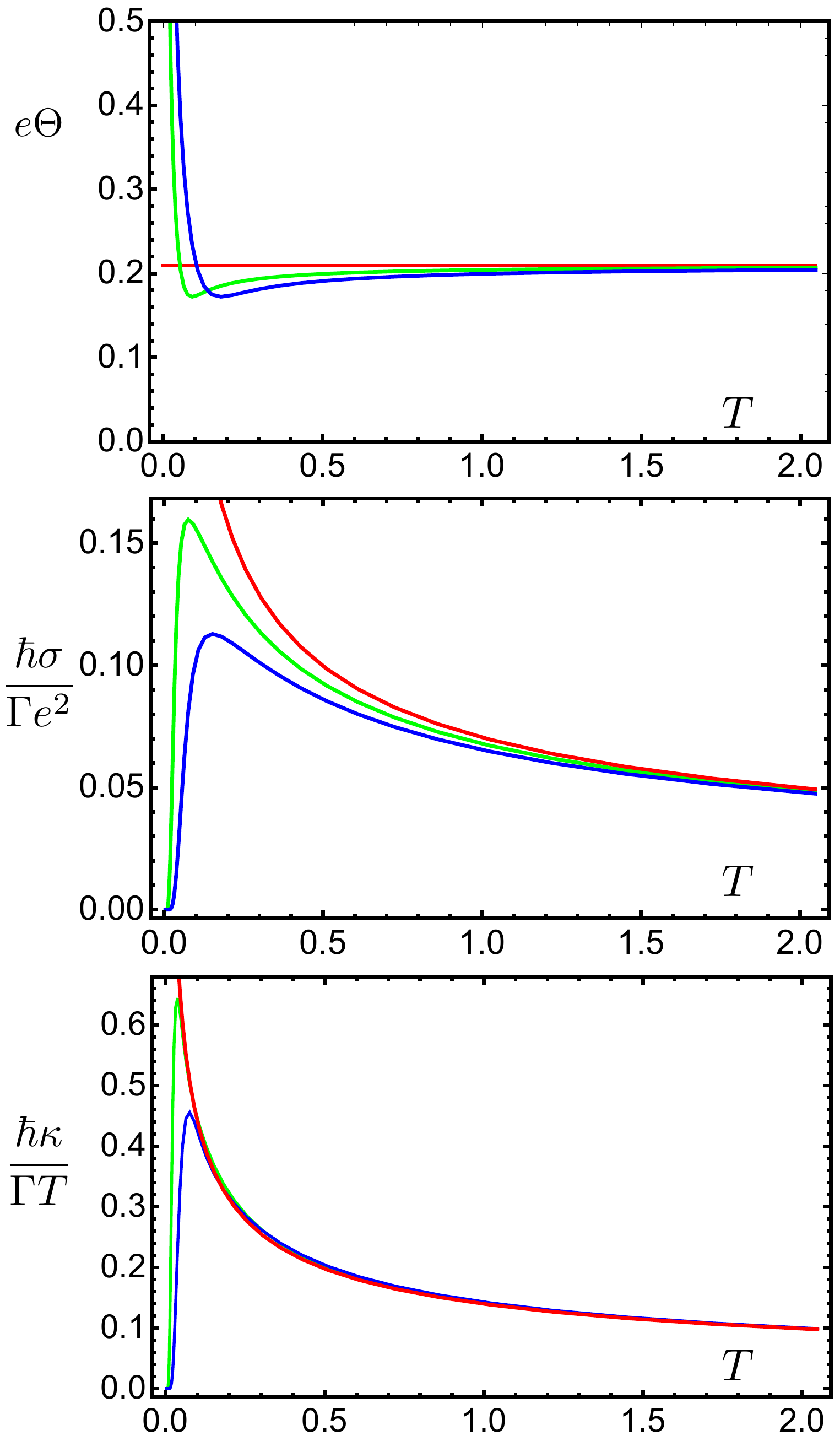} }
\caption{Crossovers in transport from the SYK regime at $T > E_{C}$, to Coulomb blockade behavior at $T < E_{C}$. It is assumed here that $T, E_{C} \gg E_{\rm coh}, J/N$. We use $J=100$ and $\mathcal{E} = 0.05$. The values of $E_C$ are $E_C = 0$ (red), $E_C = 0.2$ (green), and $E_C = 0.4$ (blue).
}
\label{fig:EC_crossover}
\end{figure}
Returning to the full Green's function in (\ref{G1}), noting that $\mathcal{G}_\lambda (\tau)$ is independent of $\varphi (\tau)$, 
we can write
\beq
\mathcal{G}(\tau) = - b^\Delta \, \mathcal{G}_\lambda (\tau) \,  \mathcal{G}_{\rm Sch} (\tau)  \label{Gfull}
\eeq
where $\mathcal{G}_\lambda (\tau)$ is specified either by (\ref{GL2}) or (\ref{GL1}), and $\mathcal{G}_{\rm Sch}$ is to be obtained from the Schwarzian path integral over $\varphi (\tau)$ which reduces to that evaluated earlier for the Majorana SYK model \cite{Altland2016,Lam:2018pvp,Yang:2018gdb,Kobrin2020}. The expression (\ref{Gfull}) is actually valid for arbitrary ratios between the energy scales $T$, $J/N$ and $E_C$. But for the case $T, E_C \gg J/N$ of interest in the present section, we can approximate the path integral over $\varphi (\tau)$ by its saddle point $\varphi(\tau) = 2 \pi \tau/\beta$, whence
\beq
\mathcal{G}_{\rm Sch} (\tau) \approx 
\left( \frac{\beta}{\pi} \sin \frac{\pi \tau}{\beta} \right)^{-2 \Delta} \,. \label{G2}
\eeq

We now take the Fourier transform of (\ref{Gfull}, \ref{G2}), and analytically continue to obtain $G( \omega)$ at real frequencies; see the Appendix of Ref.~\onlinecite{Kamenev96} for a similar computation with $n=0$ term for $\mathcal{G}_\lambda$ in (\ref{GL2}) at $\calE = 0$. This computation is described in Appendix \ref{AC}. Inserting the Green's function so obtained, we can compute the transport properties at non-zero $E_C$ from the expressions in Section~\ref{sec:setup}: the numerical results are presented in Fig.~\ref{fig:EC_crossover}.

For $T \gg E_C$, the results are described in Section~\ref{sec:SYK}. For $E_{\rm coh}, J/N \ll T \ll E_C$ we obtain the limiting forms
\bea
&& e \Theta  \sim   \mathcal{E} \frac{E_C}{T} \nn
&&\frac{\hbar \sigma}{\Gamma e^2},~\frac{\hbar \kappa}{\Gamma T}  \sim  e^{-E_C/T} \nonumber \\
&&\frac{e^2\kappa}{\sigma T}  \sim (0.13+\mathcal{O}(\mathcal{E}^2))\frac{E_C^2}{T^2}.
\label{ECregime}
\eea
Transport is exponentially suppressed by the charging energy in this Coulomb blockade regime. Nevertheless $\Theta \sim 1/T$ from the structure of the exponential factors. Terms higher order in the coupling to the leads, $\Gamma$, can become important in this Coulomb blockade regime \cite{Altland2019}, as we will discuss briefly in Section~\ref{sec:gamma}.

\section{Schwarzian corrections}
\label{sec:schwarzian}

This section considers the case where the lower energy cutoff to SYK criticality is provided by $J/N$, the third case in Section~\ref{sec:intro}. In this case, the crossover at the scale $J/N$ is described by the Schwarzian quantum gravity theory. The other lower-cutoffs in Section~\ref{sec:intro} are presumed to be at lower energy {\it i.e.} we assume in this section that $J/N \gg E_C, E_{\rm coh}$.

The expression (\ref{Gfull}) can also be used in the present section. 
Now because we assume $T, J/N \gg E_C$ in the present section, we can evaluate the $\lambda$ integral at the saddle point level, while the Schwarzian integral over $\varphi$ has to be exactly evaluated
(note that this the complement of the method in Section~\ref{sec:charging}). 
So we can use
\beq
\mathcal{G}_\lambda (\tau) \approx  \exp \left[ \calE \left( \pi - \frac{2 \pi}{\beta} \tau  \right) \right] \label{G2S}
\eeq
which is the $n=0$ term in (\ref{GL2}) for $\beta E_c \ll 1$. For $\mathcal{G}_{\rm Sch} (\tau)$, we use the expression  presented in Eq. (20) of the supplement on Ref.~\onlinecite{Kobrin2020}. We compute $\mathcal{G}(\omega)$ with this simplification in Appendix \ref{AC}. We also further compute $\mathcal{G}(\omega)$ without either of the simplifications of (\ref{G2}, \ref{G2S}) in that Appendix. 

\begin{figure}
\center{
\includegraphics[width = 2.9in]{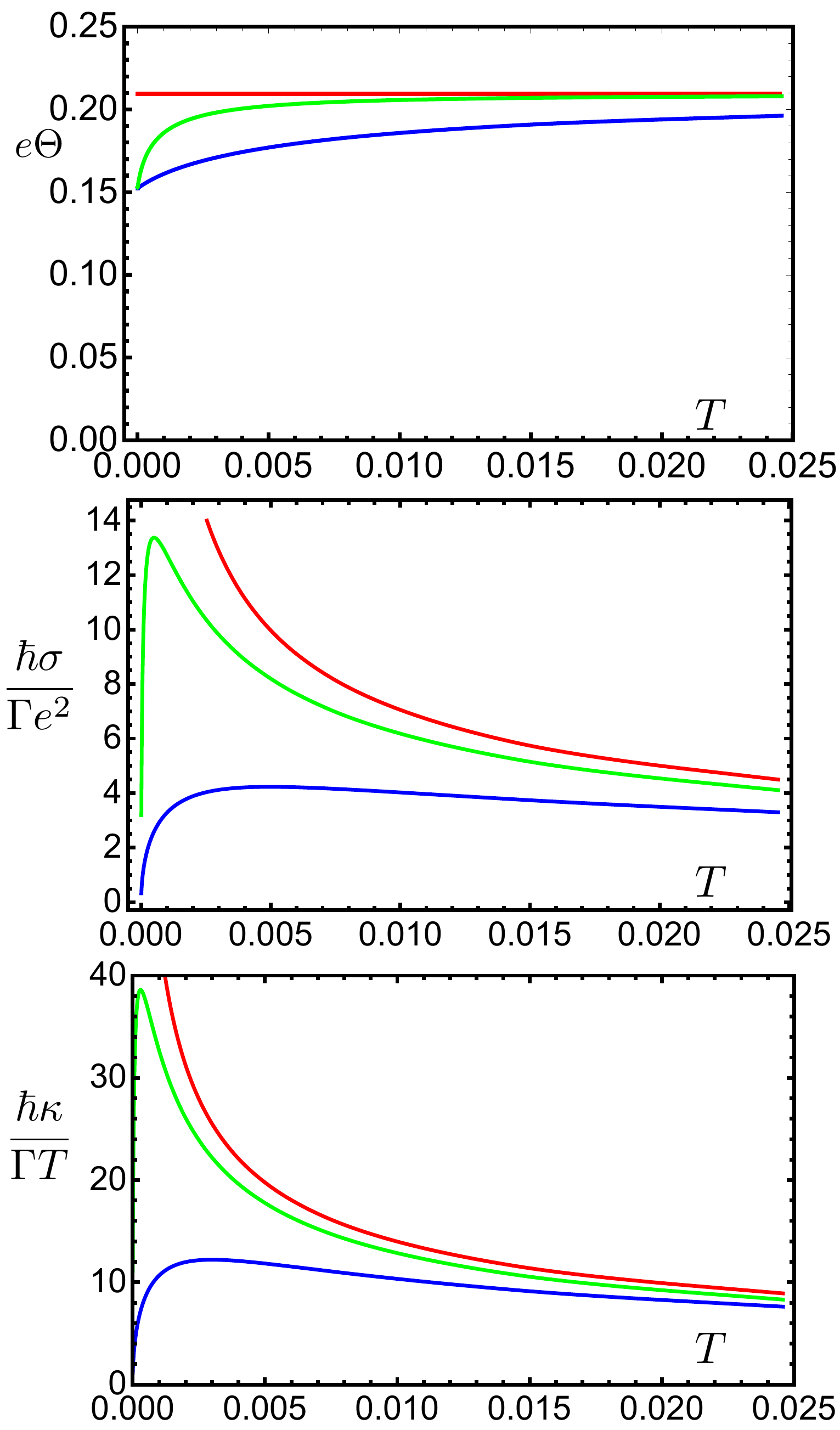} }
\caption{Crossovers in transport from the SYK regime at $T > J/N$, to Schwarizan dominated behavior at $T < J/N$. It is assumed here that $T, J/N \gg E_{\rm coh}, E_C$. We use $J=1$ and $\mathcal{E} = 0.05$. The values of $N$ are $N = \infty$ (red), $N=1000$ (green), and $N=100$ (blue).
}
\label{fig:Sch_crossover}
\end{figure}
From the $\mathcal{G} (\omega)$ so obtained, we can compute the transport properties as in Section~\ref{sec:setup}, leading to the numerical results presented in Fig.~\ref{fig:Sch_crossover}.
For $T \gg J/N$, the results are described in Section~\ref{sec:SYK}. For $E_{\rm coh}, E_C \ll T \ll J/N$ we obtain the limiting forms
\bea
e \Theta & \sim &  3.1 \mathcal{E} \nn
\frac{\hbar \sigma}{\Gamma e^2} & \sim & [1.3 + \mathcal{O}(\mathcal{E}^2)] \frac{N T^{1/2}}{J^{3/2}} \nonumber \\
\frac{\hbar \kappa}{\Gamma T} & \sim & [4.5 + \mathcal{O}(\mathcal{E}^2)] \frac{N T^{1/2}}{J^{3/2}}  \, .
\label{Schwarzianregime}
\eea
In this regime with $T \ll J/N$ there is a Schwarzian criticality with a fermion scaling dimension $\Delta =3/4$ \cite{Altland2016}, and we understand the power-laws in (\ref{Schwarzianregime}) from this value.

\section{Experimental parameters}
\label{sec:expts}

We now present some estimates of parameters for the proposed realization of SYK phases in an irregular-shaped graphene flake \cite{Chen2018}. Their proposal depends  on the physics of Dirac electrons in strong magnetic fields. 
The magnetic fields should be sufficiently strong to ensure the well-defined Landau levels in a sub-micrometer  graphene flake. One can work with $B \approx  10$ T in the current state of quantum transport experiments.  The typical length scale set by external magnetic fields is $l_B = \sqrt{h/e B}$, which for $B=$10 T gives magnetic length scale $l_B \approx 8 $ nm. We propose a graphene flake of size $L \simeq 80$ nm should allow the wavefunctions to be sensitive to the disorder originating from the edges.  For such length scales, the number of states in the lowest Landau level is hundreds, $N \approx \Phi/\Phi_0 \sim 10^2$. This sets the number of states for the SYK electrons $N$. 

We have two initial energy scales: the strength of the SYK interactions in Ref.~\onlinecite{Chen2018} is estimated as $J \approx 25$ meV, but this is likely to decrease with the effective system size $L/l_B$ ($J$ vanishes for infinite-size monolayer); we take here $J \approx 10$ meV as a reasonable estimate for our $L \simeq 80$ nm flake. The finite bandwidth of the lowest Landau level (LLL)  is set by bulk disorder and is  estimated to be $t \approx 1$ meV in realistic samples. 

These estimates yield a rather low energy for the coherence scale $E_{\rm coh} \approx 1$~K for the crossover in Section~\ref{crossoverFL}. The crossover for the Schwarzian corrections in Section~\ref{sec:schwarzian} is similar $J/N \approx 1$~K.
Finally, the charging energy $E_C$, relevant for Section~\ref{sec:charging} can be controlled by tuning the capacitance of the system independently from other effective parameters (flake size, dielectric constant, separation to the gate, and etc.). This can be achieved, for example, by putting the back gate in the proximity of the dots, which increases capacitance $C$ of the quantum dot in the controllable way. Depending on the thickness of the dielectric layer, the charging energy of the quantum dot $C$ can be tuned between 0.5 to 50 K.

\section{Conclusions}

The SYK model realizes a many-body state of quantum matter without quasiparticle excitaions. Its unusual properties include a non-vanishing entropy density in the zero temperature limit, and a maximal quantum Lyapunov exponent. Its low energy theory is identical to those of charged quantum black holes which also share these properties. 

One of the main points of our paper is that the large low $T$ entropy density has a direct experimental signature in the thermopower. We examined in detail the experimental configuration shown in Fig.~\ref{fig:sykleads}: with a SYK island, realized {\it e.g.\/} in a graphene flake as proposed in Ref.~\onlinecite{Chen2018}, coupled to metallic leads. We computed the characteristic signature of SYK criticality in the electrical conductance, the thermal conductance, and the thermopower, and the main results are in (\ref{SYKregime}).

Every experimental realizations will have some low temperature cutoff below which the SYK criticality described above will not hold. We consider three possibilities for this cutoff: the coherence energy $E_{\rm coh}$ in (\ref{Ec}) below which quasiparticles re-emerge (see Section~\ref{crossoverFL} and (\ref{FermiLiquidregime})), the charging energy $E_C$ (see Section~\ref{sec:charging} and (\ref{ECregime})), and finite $N$ effects which appear below the scale $J/N$ and are described by quantum gravity fluctuations controlled by a Schwarzian action (see Section~\ref{sec:schwarzian} and (\ref{Schwarzianregime})). The effects of these crossovers in the transport properties are contained in Figs.~\ref{fig:FL_crossover}, \ref{fig:EC_crossover}, and \ref{fig:Sch_crossover} respectively.

An examination of these figures show that the thermopower, $\Theta$, has distinctive signatures, which will allow identification of the SYK regime, and the nature of the low $T$ crossover. The thermopower is $T$ independent in the SYK regime $T<J$, and remains nearly so into the crossover into the Schwarzian dominated regime at $T < J/N$
(see Fig.~\ref{fig:Sch_crossover}); this reflects the presence of the zero temperature entropy $S_0$. For the case of a low $T$ crossover into the Fermi liquid, $\Theta$ vanishes linearly with $T$, reflecting the vanishing entropy of the Fermi liquid (see Fig.~\ref{fig:FL_crossover}); and for a low $T$ crossover into the Coulomb blockade, $\Theta$ increases (see Fig.~\ref{fig:EC_crossover}). Figs.~\ref{fig:FL_crossover}, \ref{fig:EC_crossover}, and \ref{fig:Sch_crossover} also contain the crossovers in the electrical and thermal conductance, which should serve as additional experimental diagnostics, and allow conclusive identification of a regime of a SYK criticality.

\subsection{Larger $\Gamma$}
\label{sec:gamma}

An important topic for future work is the influence of larger values of the coupling, $\Gamma$, between the SYK island and the metallic leads. It is not difficut to determine from perturbation theory that the energy scale $\Gamma^2/J$ controls the low temperature crossover out of the SYK regime due to the coupling to the leads. This is similar to coherence scale $E_{\rm coh}= t^2/J$ in (\ref{Ec}) for the influence of single particle hopping within the SYK island. Indeed, we expect the influence of the energy scale $\Gamma^2/J$ on the transport properties to be similar to the influence of $E_{\rm coh}$ in Section~\ref{crossoverFL}, as both are associated with an increase in single-particle hopping. Ref.~\onlinecite{Gnezdilov2018} considered large $\Gamma$ corrections to expressions like (\ref{dIdV}); however, their analysis does not account for the important self-consistent renormalization of the SYK Green's function that is induced by a non-zero $\Gamma$ (similar to the renormalization induced by a non-zero $t$, that we computed in Section~\ref{crossoverFL}).

Another larger $\Gamma$ effect, is the importance of inelastic co-tunneling at temperatures below the charging energy $E_C$, where the direct tunnelling contributions (computed in the present paper) are exponentially small. This was pointed out in Ref.~\onlinecite{Altland2019}, and it would be interesting to extend their analysis to the particle-hole asymmetric case with a non-zero thermopower.

\section*{Acknowledgements}

We thank  A.~Altland, L.~Anderson, D.~Bagrets, B.~Halperin, A.~Kamenev,  A.~Laitinen, and Norman Yao for enlightening discussions.  This research project was support by the U.S. Department of Energy under Grant DE-SC0019030, and by the MURI grant W911NF-14-1-0003 from ARO. AAP was supported by the Miller Institute for Basic Research in Science.

\appendix

\begin{widetext}

\section{Corrected real frequency Green's functions}
\label{AC}

In this Appendix we describe the computation of $\mathcal{G}(\omega)$ taking into account corrections due to both the $\lambda (\tau)$ fluctuations at nonzero charging energy $E_C$, and the Schwarzian fluctuations of $\varphi (\tau)$. 

We first deal with the regime appropriate for (\ref{G2}) in Section~\ref{sec:charging}, with only the effects of $\lambda (\tau)$ fluctuations with $E_C\neq0$ taken into account. Substituting (\ref{G2}, \ref{GL1}) in (\ref{Gfull}), we straightforwardly obtain, after a Fourier transform and analytic continuation
\beq
\mathcal{G}(\omega) = -\frac{e^{i\pi/4}\mathrm{sech}^{1/4}(2\pi\mathcal{E})}{2\pi^{1/4}\sqrt{T J}\vartheta_3\left(i\pi\mathcal{E},e^{-\frac{E_C}{2T}}\right)}
\sum_{p=-\infty}^\infty \frac{\left(e^{\frac{E_C(2p+1)}{2 T}}+i\right) e^{\pi \mathcal{E} (2 p+1)-\frac{E_C (p+1)^2}{2 T}} \Gamma \left(\frac{E_C i (2 p+1)}{4 \pi  T}+\frac{1}{4}-\frac{i \omega }{2 \pi  T}\right)}{\Gamma \left(\frac{E_C i (2 p+1)}{4 \pi  T}+\frac{3}{4}-\frac{i \omega }{2 \pi  T}\right)},
\label{GR2}
\eeq
where $\vartheta_3$ is a Jacobi theta function \cite{varthetadefn}. The sum over $p$ must be evaluated numerically but converges rapidly.

If we instead go to the regime appropriate for (\ref{G2S}) in Section~\ref{sec:schwarzian} which includes only flucuations of $\varphi (\tau)$, we must use the expression presented in Eq. (20) of the supplement on Ref.~\onlinecite{Kobrin2020} for $\mathcal{G}_{\rm Sch} (\tau)$, which is 
\bea
&&\mathcal{G}_\mathrm{Sch}(\tau) = \frac{e^{-2 \pi ^2 C_\mathrm{Sch} T}}{C_\mathrm{Sch}^2 J^{1/2} T^{3/2}}\int_0^\infty \frac{ds_1}{2\pi^2}\int_0^\infty \frac{ds_2}{2\pi^2} \Bigg[ s_1 s_2 \sinh(2\pi s_1)\sinh(2\pi s_2) \nn
&&\times e^{-\frac{\tau  T (s_1^2-s_2^2)+s_2^2}{2 C_\mathrm{Sch} T}}\left| \Gamma \left(\frac{1}{4}-i (s_1+s_2)\right) \Gamma \left(\frac{1}{4}+i (s_1-s_2)\right)\right| ^2 \Bigg],
\eea
along with (\ref{G2S}) for $\mathcal{G}_\lambda$. We do the imaginary-time Fourier transformation first. This involves the integral
\beq
\int_0^{1/T}d\tau~e^{i\omega_n\tau} e^{-\frac{s_1^2\tau+s_2^2(1/T-\tau)}{2C_\mathrm{Sch}}}e^{\pi\mathcal{E}(1 -2 T\tau)} 
= \frac{2 C_\mathrm{Sch} e^{-\pi\mathcal{E}} \left(e^{-\frac{s_1^2}{2 C_\mathrm{Sch} T}}+e^{2 \pi\mathcal{E}-\frac{s_2^2}{2 C_\mathrm{Sch} T}}\right)}{4 \pi  C_\mathrm{Sch} \mathcal{E} T-2 i C_\mathrm{Sch} \omega_n+s_1^2-s_2^2},
\eeq
where we used $e^{i(\omega_n/T-\pi)}=1$. For numerical purposes it is easier to analytically continue $i\omega_n \rightarrow \omega + i0^+$, and then take the imaginary part to compute the spectral function $\mathcal{A}(\omega) = -(1/\pi)\mathrm{Im}[\mathcal{G}(\omega)]$, 
\beq
\frac{2 C_\mathrm{Sch} e^{-\pi\mathcal{E}} \left(e^{-\frac{s_1^2}{2 C_\mathrm{Sch} T}}+e^{2 \pi\mathcal{E}-\frac{s_2^2}{2 C_\mathrm{Sch} T}}\right)}{4 \pi  C_\mathrm{Sch} \mathcal{E} T-2 i C_\mathrm{Sch} \omega_n+s_1^2-s_2^2} \rightarrow - 2 C_\mathrm{Sch} e^{-\pi\mathcal{E}} \left(e^{-\frac{s_1^2}{2 C_\mathrm{Sch} T}}+e^{2 \pi\mathcal{E}-\frac{s_2^2}{2 C_\mathrm{Sch} T}}\right) \delta \left(4 \pi  C_\mathrm{Sch} \mathcal{E} T-2 C_\mathrm{Sch} \omega+s_1^2-s_2^2\right).
\eeq
This $\delta$ function then leads to the final expression with only one numerical integration,
\bea
&&\mathcal{A}(\omega) = \frac{2 e^{-\pi(2\pi C_\mathrm{Sch} T+\mathcal{E})}\mathrm{sech}^{1/4}(2\pi\mathcal{E})}{C_\mathrm{Sch} J^{1/2} T^{3/2}}\left(e^{\omega /T}+1\right) \int_0^\infty~\frac{ds_1}{8\pi^4}\Bigg[\theta\left(s_1^2+4\pi C_\mathrm{Sch} T\mathcal{E}-2C_\mathrm{Sch}\omega \right)s_1\sinh(2\pi s_1)e^{-\frac{s_1^2}{2 C_\mathrm{Sch} T}} \nn
&&\times\sinh\left(2\pi\sqrt{s_1^2+4\pi C_\mathrm{Sch}T\mathcal{E}-2C_\mathrm{Sch}\omega}\right) \left| \Gamma \left(\frac{1}{4}-i \left(s_1+\sqrt{s_1^2+4 C_\mathrm{Sch} \pi  T \mathcal{E}-2 C_\mathrm{Sch} \omega }\right)\right)\right|^2 \nn
&&\times \left|\Gamma \left(\frac{1}{4} + i \left(s_1-\sqrt{s_1^2+4 C_\mathrm{Sch} \pi  T \mathcal{E}-2 C_\mathrm{Sch} \omega }\right)\right)\right| ^2\Bigg],
\label{GR2S}
\eea
where $\theta$ is the Heaviside step function. The real part can then be obtained by numerically performing the Hilbert transform
\beq
\mathrm{Re}[\mathcal{G}(\omega)] = \fint_{-\infty}^{\infty}d\Omega~\frac{\mathcal{A}(\Omega)}{\omega-\Omega}.
\label{ReGR}
\eeq

Finally, we note the computation for the full Green's function (\ref{Gfull}) using Eq. (20) of the supplement on Ref.~\onlinecite{Kobrin2020} for $\mathcal{G}_{\rm Sch} (\tau)$, along with (\ref{GL1}) for $\mathcal{G}_\lambda$. The following result is valid for arbitrary ratios between $\omega$, $T$, $E_C$ and $J/N$, provided they are all smaller than $J$. It is similar to the above, but with an additional numerical summation, and the final result is,
\bea
&&\mathcal{A}(\omega) = \frac{2e^{-2 \pi ^2 C_\mathrm{Sch} T} \mathrm{sech}^{1/4}(2 \pi \mathcal{E})}{C_\mathrm{Sch} J^{1/2} T^{3/2}}\frac{\left(e^{\omega/T}+1\right)}{\vartheta_3\left(i\pi\mathcal{E},e^{-\frac{E_C}{2T}}\right)}\sum_{p=-\infty}^\infty \int_0^\infty~\frac{ds_1}{8\pi^4}\Bigg[\theta\left(C_\mathrm{Sch} (2 E_C p+E_C-2 \omega )+s_1^2\right)s_1\sinh(2\pi s_1) \nn
&&\times \exp \left(-\frac{C_\mathrm{Sch} E_C (p+1)^2-2 \pi  C_\mathrm{Sch} \mathcal{E} (2 p+1) T+s_1^2}{2 C_\mathrm{Sch} T}\right)\sinh \left(2 \pi  \sqrt{C_\mathrm{Sch} (2 E_C p+E_C-2 \omega )+s_1^2}\right) \nn
&&\times \left| \Gamma \left(\frac{1}{4}-i \left(s_1+\sqrt{s_1^2+C_\mathrm{Sch} (2 p E_C+E_C-2 \omega )}\right)\right) \Gamma \left(\frac{1}{4}+ i \left(s_1-\sqrt{s_1^2+C_\mathrm{Sch} (2 p E_C+E_C-2 \omega )}\right)\right)\right| ^2,
\label{GRfull}
\eea
with the real part again determined as in the above. The numerical integrations and summations can be peformed using \verb|NIntegrate|, \verb|Sum| for (\ref{GR2}) (with a cutoff on $p$), and \verb|NSum| for (\ref{GRfull}) in \verb|Mathematica| without any issues. For (\ref{ReGR}), the numerical integration can be done with \verb|NIntegrate| using the standard definition of the Cauchy principal value.

\end{widetext}

\bibliography{Refs}

\end{document}